\documentclass{article}%
\usepackage{amsmath}
\usepackage{amsfonts}
\usepackage{amssymb}
\usepackage{graphicx}%
\begin{document}

\author{Bruno COURCELLE\\LaBRI, CNRS,\\351\ Cours de la Lib\'{e}ration,\\33405\ Talence, France\\courcell@labri.fr}
\title{Clique-width and edge contraction}
\maketitle

\begin{quote}
\textbf{Abstract} : We prove that edge contractions do not preserve the
property that a set of graphs has bounded clique-width.\ 

\end{quote}

\bigskip

\bigskip

\section{\bigskip Introduction}

Clique-width is, like tree-width, an integer graph invariant that is an
appropriate parameter for the contruction of many fixed-parameter
tractable\ algorithms ([4, 6, 7, 11]). It is thus important to know that the
graphs of a particular type have bounded tree-width or clique-width. The
article [13] is a survey of clique-width bounded classes. Gurski has reviewed
in [9] how clique-width behaves under different graph operations. He asks
whether, for each $k$, the class of graphs of clique-width at most $k$ is
stable under edge contractions.\ This is true for $k=2$, \emph{i.e.}, for
cographs, and we prove that this is false for $k=3$.\ (For each $k$, this
stability property is true for graphs of tree-width at most $k$.\ It is thus
natural to ask the question for clique-width.)

Gurski proves that contracting one edge can at most double the
clique-width.\ The conjecture is made in [14] (Conjecture 8) that contracting
several edges in a graph of clique-width $k$ yields a graph of clique-width at
most $f(k)$ for some fixed function $f$. We disprove this conjecture and
answer Gurski's question by proving the following proposition.

\bigskip

\textbf{Proposition 1} : The graphs obtained by edge contractions from graphs
of clique-width 3 or of linear clique-width 4, have unbounded clique-width.

\bigskip

The graphs of clique-width at most 2 (they are the cographs) are preserved
under edge contractions.The validity of the conjecture of [14] would have
implied that the \emph{restricted vertex multicut problem} is fixed-parameter
tractable\ if the parameter is the clique-width of a certain graph describing
the input in a natural way.\ This problem consists in finding a set of
vertices of given size that meets every path between the two vertices of each
pair of a given set and does not contain any vertex of these pairs. Without
Conjecture 8, this problem is fixed-parameter tractable under the additional
condition that no two vertices from different pairs are adjacent.\ 

\bigskip

For sake of comparison, we also consider contractions of edges, one end of
which has degree 2. We say in this case that we \emph{erase a vertex}: we
erase $x$ if it has exactly two neighbours ; to do so, we add an edge between
them (unless they are adjacent, we only consider graphs without parallel
edges) and we delete $x$ and its two incident edges. The graphs obtained from
a graph by erasing and deleting vertices are its \emph{induced topological
minors}.

\bigskip

\textbf{Proposition 2:}\ The induced topological minors of the graphs of
clique-width $k$ have clique-width at most $2^{k+1}-1$.

\bigskip

\section{Some basic facts}

\bigskip

Graphs are finite, undirected, loop-free and without parallel edges. To keep
this note as short as possible, we refer the reader to any of [3, 11, 13, 15,
16] for the definitions of \emph{clique-width }and\emph{ rank-width}.\ Other
references for clique-width are [1, 5, 8, 14]. A variant of clique-width
called \emph{linear clique-width} is studied in [3, 10]. We denote by $cwd(G)$
and $rwd(G)$ the clique-width and, respectively, the rank-width of a graph
$G$.\ We recall from [16] that we have $rwd(G)\leq cwd(G)\leq2^{rwd(G)+1}-1$.
Proving that $cwd(G)>k$ for given $G$ and $k$ is rather difficult in most
cases.\ (See for instance the computation of the exact clique-width of a
square grid in [8].\ The computation of its rank-width in [12] is not
easier.)\ We overcome this difficulty by using \emph{monadic second-order
transductions} : they are graph transformations specified by formulas of
monadic second-order logic. The (technical) definition is in [2,3].\ We will
only need the fact that the graphs defined by a monadic second-order
transduction $\tau$\ from graphs of clique-width at most $k$ have clique-width
at most $f_{\tau}(k)$ for some computable function $f_{\tau}$ that can be
determined from the formulas forming the definition of $\tau$ (Corollary
7.38(2), [3]). However, we also give an alternative proof based on rank-width
and vertex-minors that does not use monadic second-order transductions.

\bigskip

A \emph{vertex-minor} of a graph is obtained by deleting vertices\ (and the
incident edges) and performing \emph{local complementations}. (Local
complementation exchanges edges and non-edges between the neighbours of a
vertex.)\ These transformations do not increase rank-width [15].\ Erasing\ a
vertex $x$ yields a vertex-minor of the considered graph: let $y$ and $z$ be
its neighbours; if they are adjacent, erasing $x$ is the same as deleting it
because we fuse parallel edges; if they are not, erasing $x$ is the same as
performing a local complementation at $x$ (which creates an edge between $y$
and $z$), and then deleting $x$.\ Hence, by transitivity, every induced
topological minor is a vertex-minor.

\section{Proofs}

\bigskip

\bigskip\textbf{Definitions and notation }

(a) We denote by $H/F$\ the graph obtained from a graph $H$ by contracting the
edges of a set $F$. (Parallel edges are fused, no loops are created.) If
$\mathcal{H}$ is a set of graphs, we denote by $EC(\mathcal{H})$ the set of
graphs $H/F$ such that $H\in\mathcal{H}$ and $F$ is a set of edges of $H$.\ 

(b) We denote by $\mathcal{R}$ the set of graphs having a \emph{proper edge
coloring} with colors in $\{1,...,4\}$: every two adjacent edges have
different colors.\ These graphs have unbounded tree-width and clique-width as
they include the square grids (the $n\times n$ grid has clique-width $n+1$ if
$n\geq2$ by [8]).

(c) For $n\geq2,$ we define a graph $G_{n}$.\ Its vertices are $x_{1}%
,...,x_{n},y_{1},...,y_{n}$ \ and its edges are $x_{i}-y_{i},y_{i}-y_{j}$ for
all $i,j\neq i.$ (The notation $x-y$ designates an edge between $x$ and $y$).
We let $D$ consist of 4 vertices and no edge, and we let $H_{n}$ be obtained
from $G_{n}$ by substituting disjoint copies of $D$ to each vertex $y_{i}%
$.\ Hence and more precisely, $H_{n}$ has the $5n$ vertices $x_{1}%
,...,x_{n},y_{1}^{1},y_{1}^{2},y_{1}^{3},y_{1}^{4},y_{2}^{1},...,y_{n}^{4}$
and the $8n^{2}-4n$ edges $x_{i}-y_{i}^{c},y_{i}^{c}-y_{j}^{d}$ for all
$i,j\neq i$ and $c,d=1,...,4$. We denote by $\mathcal{H}$ the set of graphs
$H_{n}$. It is easy to construct expressions of $G_{n}$ and $H_{n}$ showing
that they have clique-width at most 3 and linear clique-width at most 4. If
$n\geq3,$ they have clique-width 3 because they contain, as an induced
subgraph, the path with 4 vertices, so that they do not have clique-width 2,
and linear clique-width 4 because they contain, as an induced subgraph, the
graph $G_{3}$ that is not a cocomparability graph, so that they do not have
linear clique-width at most 3 by Proposition 14\ of [10].

(d) We define a monadic second-order transduction $\alpha$ with one parameter
$X$.\ If $G$ is a graph and $X$ is a set of vertices, then the graph
$\alpha(G,X)$ is defined if $X$ is stable (no two vertices are adjacent); its
vertex set is $X$ and it has an edge between $x$ and $y$ if and only if these
vertices are at distance 2 in $G$.\ We denote by $\alpha(G)$ the set of all
such graphs, and by $\alpha(\mathcal{G)}$ the union of the sets $\alpha(G)$
for $G$ in a set $\mathcal{G}$.

\bigskip

\textbf{Lemma 3 }: We have $\alpha(EC(\mathcal{H}))\supseteq\mathcal{R}.$\ 

\bigskip

\textbf{Proof}: Let $R$ be a graph in $\mathcal{R}$ with vertices
$x_{1},...,x_{n}$ and a proper edge coloring with colors 1 to 4.\ The set
$X=\{x_{1},...,x_{n}\}$ is also a subset of the vertex set of $H_{n}$. The
four neighbours of $x_{i}$ in $H_{n}$ are $y_{i}^{1},y_{i}^{2},y_{i}^{3}$
\ and $y_{i}^{4}.$

Let $F$ be the set of edges of the form $y_{i}^{c}-y_{j}^{c}$ such that
$x_{i}-x_{j}$ is an edge of $R$ colored by $c$, $c\in\{1,...,4\}$
(hence,\ $x_{i}$ and $x_{j}$ are at distance 3 in $H_{n}$). The graph
$K=H_{n}/F$ belongs to $EC(\mathcal{H})$ and $X$ is stable in this graph (the
vertices $x_{1},...,x_{n}$ are not affected by the contractions of edges). It
is clear that $x_{i}-x_{j}$ is an edge of $R$ if and only if there is in $K$ a
path\ $x_{i}-z-x_{j}$ where $z$ results from the contraction of the edge
$y_{i}^{c}-y_{j}^{c}$ and $c$ is the color of $x_{i}-x_{j}$ in $R$.\ It
follows that $R=\alpha(K,X)$. $\square$

\bigskip

\textbf{Proof of Proposition 1: }By Lemma 3, the set $\alpha(EC(\mathcal{H}))$
has unbounded clique-width.\ Hence, so has $EC(\mathcal{H})$ by Corollary
7.38(2) of [3] recalled above. This concludes the proof because the graphs
$H_{n}$ have clique-width 3 and linear clique-width 4 for $n\geq3$. $\square$

\bigskip

\emph{NLC-width} and clique-width are linearly related (see [9]). Hence, the
graphs obtained by edge contractions from graphs of NLC-width at most 3 have
unbounded NLC-width.\ 

\textbf{Remark: }For each $n,$ the graph $H_{m^{2}}$ of clique-width 3 having
$5m^{2}$\ vertices where $m=f_{\alpha}(n)$ yields by edge contractions a graph
of clique-width at least $n+1$.\ Here, $f_{\alpha}$ is the computable function
of Section 2 that can be assumed monotone.\ To prove this, we let $F$ be a set
of edges such that $\alpha(H_{m^{2}}/F)$ contains the $m\times m$ grid $R_{m}$
and $k=cwd(H_{m^{2}}/F)$. Then, $m+1=cwd(R_{m})\leq f_{\alpha}(k)$, hence
$f_{\alpha}(n)+1\leq f_{\alpha}(k),$\ and so, $k>n$.\ The function $f_{\alpha
}$ is very fast growing.\ A much better upper-bound will be obtained from the
alternative proof we give next.

\bigskip

Edge contractions can increase rank-width because the same sets of graphs have
bounded rank-width and bounded clique-width [16]. The following proof shows
this directly.

\bigskip

\textbf{Alternative proof of Proposition 1}\ : The construction is similar and
we use the same notation. We construct $H_{n}^{\prime}$ from $G_{n}$ by
substituting disjoint copies of $K_{4}$ to each vertex $y_{i}$ and by adding a
vertex $y_{0}$ adjacent to all vertices $y_{1}^{1},y_{1}^{2},y_{1}^{3}%
,y_{1}^{4},y_{2}^{1},...,y_{n}^{4}$. Hence, $H_{n}^{\prime}$ has $5n+1$
vertices.We denote by $\mathcal{H}^{\prime}$ the set of graphs $H_{n}^{\prime
}$. They have clique-width 3 and linear clique-width 4 (by the same argument
as for $H_{n}$).

Let us fix $n$ and let $R$ be the $n\times n$ grid with vertices
$x_{1},...,x_{n^{2}}$. To prove that it is a vertex-minor of $H_{n^{2}%
}^{\prime}$, we take a proper edge-coloring of $R$ with colors $1,...,4$, we
contract the edges $y_{i}^{c}-y_{j}^{c}$ of $H_{n^{2}}^{\prime}$ such that
$x_{i}-x_{j}$ is an edge of $R$ colored by $c$.\ This gives a graph
$R^{\prime}$ that has $R$ as vertex-minor.\ To prove this, we delete the
vertices $y_{i}^{c}$ such that $x_{i}$ has no incident edge colored by $c$, we
take a local complementation at $y_{0}$, we delete $y_{0}$ and finally, we
erase the vertices resulting from the contraction of the edges $y_{i}%
^{c}-y_{j}^{c}$ after taking local complementations at them.

The rank-width of $R$ is $n-1$ by [12], that of $R^{\prime}$ is thus at least
$n-1,$ and so is its clique-width. Hence, by contracting edges in a graph of
clique-width 3 (and linear clique-width 4) that has $5n^{2}+1$ vertices, one
can get a graph of clique-width at least $n-1$.$\square$

\bigskip

\textbf{Remark:} An algorithm can determine a graph of clique-width 3 that
yields a graph of clique-width more than 3 by the contraction of a single
edge. It performs an exhaustive search until some graph is obtained: for each
$n=2,3$ ... it considers the finitely many sets $F$ of\ pairwise nonadjacent
edges of $H_{n}$.\ By using the polynomial-time algorithm of [1] to check if a
graph has clique-width at most 3, it can look for a set\ $F$ and an edge $f\in
F$ such that $H_{n}/(F-\{f\})$ has clique-width 3 and $H_{n}/F$ has
clique-width more than 3 (actually 4, 5 or 6 by Theorem 4.8 of [9]). By
Proposition 1, one must find some $n$ and such $F$ and $f$. However, we have
not implemented this algorithm.

\bigskip

Gurski has proved that erasing a\ vertex of degree 2 can increase (or
decrease) the clique-width by at most 2.\ In Proposition 2, we consider the
effect of erasing several vertices and taking induced subgraphs.\ 

\bigskip

\textbf{Proof of Proposition 2}\ : As noted above, an induced topological
minor is a vertex-minor. The result follows since, for every graph $G$, we
have $rwd(G)\leq cwd(G)\leq2^{rwd(G)+1}-1$.$\square$

\bigskip

This proof leaves open the question of improving the upper bound\ $2^{k+1}-1$,
possibly to a polynomial in $k$ or even to $k$.\ 

\bigskip

\textbf{Acknowledgement:} I thank the anonymous referee who suggested the
alternative proof of Proposition 1, and M.\ Kant\'{e} and D.\ Meister for
useful comments.

\bigskip

{\large References}

\bigskip

[1] D.\ Corneil, M. Habib, J.-M. Lanlignel, B. Reed and U. Rotics:
Polynomial-time recognition of clique-width at most 3 graphs. \emph{Discrete
Applied Mathematics} \textbf{160} (2012) 834-865.

\bigskip

[2] B.\ Courcelle, Monadic Second-Order Definable Graph Transductions: A
Survey. \emph{Theor. Comput. Sci.} \textbf{126} (1994) 53-75.

\bigskip

[3] \ B. Courcelle and J. Engelfriet, \emph{Graph structure and monadic
second-order logic, a language theoretic approach}, Cambridge University
Press, 2012.

\bigskip

[4] B. Courcelle, J. Makowsky and U. Rotics, Linear Time Solvable Optimization
Problems on Graphs of Bounded Clique-Width. \emph{Theory Comput. Syst.}
\textbf{33} (2000) 125-150.

\bigskip

[5] B. Courcelle and S. Olariu, Upper bounds to the clique width of graphs.
\emph{Discrete Applied Mathematics }\textbf{101 }(2000) 77-114.

\bigskip

[6] R.\ Downey and M.\ Fellows, \emph{Parameterized Complexity}, Springer, 1999.

\bigskip

[7] J.\ Flum and M. Grohe, \ \emph{Parameterized Complexity Theory}, Springer, 2006.

\bigskip

[8] \ M.\ Golumbic and U.\ Rotics: On the Clique-Width of Some Perfect Graph
Classes. \emph{Int. J. Found. Comput. Sci. }11 (2000) 423-443.

\bigskip

[9] F.\ Gurski, Graph operations on clique-width bounded graphs, 2007, CoRR abs/cs/0701185.

\bigskip

[10] P. Heggernes, D. Meister and C. Papadopoulos, Graphs of linear
clique-width at most 3, \emph{Theoretical Computer Science} \textbf{412}
(2011) 5466-5486.

\bigskip

[11] P.\ Hlin\v{e}n\'y, S. Oum, D.\ Seese and G. Gottlob: Width Parameters
Beyond Tree-width and their Applications. \emph{Comput. J}. \textbf{51 }(2008) 326-362.

\bigskip

[12] V. Jel\'{\i}nek, The rank-width of the square grid, \emph{Discrete
Applied Mathematics} \textbf{158} (2010) 841-850.

\bigskip

[13] \ M.\ Kaminski, V. Lozin and M. Milanic: Recent developments on graphs of
bounded clique-width. \emph{Discrete Applied Mathematics} \textbf{157} (2009) 2747-2761.

\bigskip

[14] M.\ Lackner, R. Pichler, S. R\"{u}mmele and S. Woltran: Multicut on
Graphs of Bounded Clique-Width, in "Proceedings of COCOA'12",
\emph{Lec.\ Notes Comp.\ Sci.}, \textbf{7402} (2012) 115-126.

\bigskip

[15] S.\ Oum: Rank-width and vertex-minors, \emph{J. Comb. Theory, Ser. B
}\textbf{95} (2005) 79-100.

\bigskip

[16] S. Oum and P. Seymour: Approximating clique-width and branch-width.
\emph{J. Comb. Theory, Ser. B} \textbf{96} (2006) 514-528.

\bigskip

\end{document}